\documentclass[namedreferences]{solarphysics}

\usepackage[hyperref,optionalrh,showbiblabels]{spr-sola-addons} 
\usepackage{graphicx}        
\usepackage{color}           
\usepackage{breakurl}        

\chardef\us=`\_

\begin{document}

\begin{article}
\begin{opening}

\title{Meteospace, a New Instrument for Solar Survey at the Calern Observatory}

\author{J.-M.~\surname{Malherbe}$^{1}$
\sep
        Th.~\surname{Corbard}$^{2}$
\sep
        K.~\surname{Dalmasse}$^{3}$
\sep
        The Meteospace team }

\runningauthor{J.-M. Malherbe \emph{et al.}}

\runningtitle{Meteospace, a New Instrument for Solar Survey}

\institute{$^{1}$ LESIA, Observatoire de Paris, 92195 Meudon, and
PSL Research University, France,
                 email: \url{Jean-Marie.Malherbe@obspm.fr}\\
              $^{2}$ Universit\'{e} C\^{o}te d'Azur, Observatoire de la C\^{o}te d'Azur,
              CNRS, Laboratoire Lagrange, France,
               email: \url{Thierry.Corbard@oca.eu}\\
              $^{3}$ IRAP, Universit\'{e} de Toulouse, CNRS, CNES, UPS, 31028 Toulouse,
              France,
               email: \url{Kevin.Dalmasse@irap.omp.eu}
               }

\begin{abstract}
High cadence observations of solar activity (active regions, flares,
filaments) in the H$\alpha$ line were performed at Meudon and Haute
Provence Observatories from 1956 to 2004. More than 7 million images
were recorded, mainly on 35 mm films. After a review of the
scientific interest of solar surveys at high temporal resolution and
the historical background, we describe the new instrument which will
operate automatically in 2020 at the Calern station of the C\^{o}te
d'Azur observatory (1270 m). It will replace the former heliographs
with improved cadence, seeing and time coverage. We summarize the
capabilities of the optical design and present new scientific
perspectives in terms of flare onset and Moreton wave detection.
\end{abstract}
\keywords{chromosphere; full-sun; flares; active regions; filaments;
Moreton waves; heliograph}
\end{opening}

\section{Introduction} \label{S-Introduction}

Systematic observations of the full solar disk started at the Meudon
Observatory in 1908 with Deslandres spectroheliograph
\citep{Malherbe}. Spectroheliograms are obtained by spectroscopic
scans of the Sun, producing line profiles and narrow bandpass images
at constant wavelength. This is a low cadence instrument dedicated
to the study of long term solar activity.

In addition several heliographs performed high cadence (60 seconds)
H$\alpha$ observations from 1956 to 2004 at Meudon and Haute
Provence (OHP) observatories, which produced more than 7 million
images. This survey of fast solar activity (flares, evolution of
active regions and filaments) was based on Lyot filters and started
in the frame of the International Geophysical Year (IGY 1957).
Unfortunately, only 10\% of the collection is digitized, the oldest
part being on films. Nowadays, such observations are mainly done
with narrow band imagers such as Fabry P\'{e}rot filters
\citep[\emph{e.g.} Global Oscillation Network Group (GONG) H$\alpha$
network:][]{Harvey}, but some Lyot filters remain in activity
\citep[\emph{e.g.} Global H$\alpha$ Network (GHN):][]{Gallagher}.

This work is intended to promote high cadence observations of the
Sun in the context of space weather monitoring. Section 2 recalls
the scientific interest of such observations. In Section 3, we
present the historical background summarizing the evolutions and
main discoveries of the Meudon and OHP heliographs (1956-2004).
Section 4 describes the new instrument which is going to replace
them at Calern observatory. Section 5 discusses new perspectives in
terms of flare onset and Moreton waves detection.


\section{Scientific Goal of High Cadence Observations} \label{S-Scientific-Context}

%
%
Solar flares and coronal mass ejections (CMEs) are energetic events
associated with the sudden release of magnetic energy in response to
the development of resistive \citep{Furth1963,Antiochos1999} or
ideal \citep{Amari1999,Kliem2006} MHD instabilities in the coronal
magnetic field. During these events, typically $10^{28} - 10^{33}$
erg of magnetic energy are released in about $10^{3} - 10^{4}$ s
\citep{Shibata2011}. The energy is released by magnetic reconnection
and converted into thermal, kinetic and radiated energy
\citep{Carmichael1964,Sturrock1966,Hirayama1974,Kopp1976}. Through
particle acceleration and ejections, flares and CMEs are the two
major drivers of space weather that can cause various environmental
hazards at the Earth \citep{Schrijver2012}, and for which
high-cadence observations of the Sun are required.

%
%
H$\alpha$ is the best line for fast imaging of chromospheric
structures such as filaments, which are cool and dense material that
appear as elongated and dark structures. They are formed of
chromospheric plasma confined in highly stressed coronal magnetic
fields that overlay polarity inversion lines
\citep{Aulanier1998,Schmieder2006}. Filaments are typical precursors
of flares, including confined flares associated with failed filament
eruption \citep{Torok2005} and flares for which the filament erupts
and leads to the formation of a CME \citep{Moore2001}. In the latter
case, the filament is usually characterized by a slow rise (taking
hours) until it reaches a critical height beyond which the system
becomes unstable and the filament erupts
\citep{Kahler1988,Sterling2005}. In this regard, high-cadence
observations of filaments are needed for near real-time prediction
of their ejection.

%
%
H$\alpha$ imaging also provides observations of flare ribbons, a
useful signature for real-time detection of flares for which there
is either no filament eruption \citep{Dalmasse2015} or no filament
at all \citep{Masson2009}. Flare ribbons are surface brightenings
that can be observed in ultraviolet (UV) and H$\alpha$
\citep{Schmieder1987}. They are caused by the interaction of
energetic particles and thermal energy (produced at the reconnection
region) with the lower and denser atmospheric layers \citep[review
by][]{Fletcher2011}. The spatio-temporal evolution of H$\alpha$
ribbons provides information on the flare dynamics. If one further
combines H$\alpha$ data with photospheric magnetograms and some sort
of modeling, it is then possible to derive the reconnection rate
\citep{Forbes1984,Qiu2002}, the energy release rate
\citep{Asai2004,Isobe2005}, and the flare energy
\citep{Toriumi2017}.

%
%
Moreton waves are another flare-related signature for which
high-cadence imaging is required. They are large-scale disturbances
propagating in the solar atmosphere in H$\alpha$ observations
\citep{Moreton1960}. They materialize as arc-shaped, bright fronts
in the center and blue wing of the H$\alpha$ line, which propagate
out to distances of up to 600 Mm in the extreme cases \citep[review
by][]{Warmuth2015}. Their typical propagation speed is in the range
400 - 2500 km s$^{-1}$ \citep{naru,muhr,Liu2013}. The speed and
signature argue in favour of a chromospheric counterpart of a
coronal perturbation that compresses and pushes the chromosphere
downwards \citep{Uchida1968}. Coronal waves were detected by the
Extreme ultraviolet Imager Telescope (EIT) onboard SOHO
\citep{Klassen} and later by the Atmospheric Imaging Assembly (AIA)
onboard SDO \citep{Nitta}, with better temporal and spatial
resolution. Their physical nature (CME-driven or freely propagating
MHD shock-wave) is not yet fully understood.

%
%
Moreton waves can disturb the ambient coronal magnetic field and
remote filaments, and seem to be associated with strongest flare or
CME events \citep{Warmuth2015}, that are more likely to be very
geo-effective. Moreton waves detection at high-cadence, by running-
or base-difference H$\alpha$ imaging of the chromosphere (Section
5), should allow to identify in real-time the potentially most
energetic and dangerous eruptive events in space weather
applications.

In this context, the new instrument presented here has two major
goals:

\begin{description}
  \item[\textnormal{i)}] solar activity monitoring in chromospheric lines
  as H$\alpha$ center (flares and filaments) and Ca\textsc{ii} K (magnetic
  proxy);
  \item[\textnormal{ii)}] flare, CME onset and Moreton wave detection using two high
  cadence (15 seconds)
  H$\alpha$ filters (center and blue wing).
\end{description}


\section{Historical Background of the French Heliographs} \label{S-Historical-Context}

\begin{figure}
\centering
\includegraphics[width=1.0\textwidth,clip=]{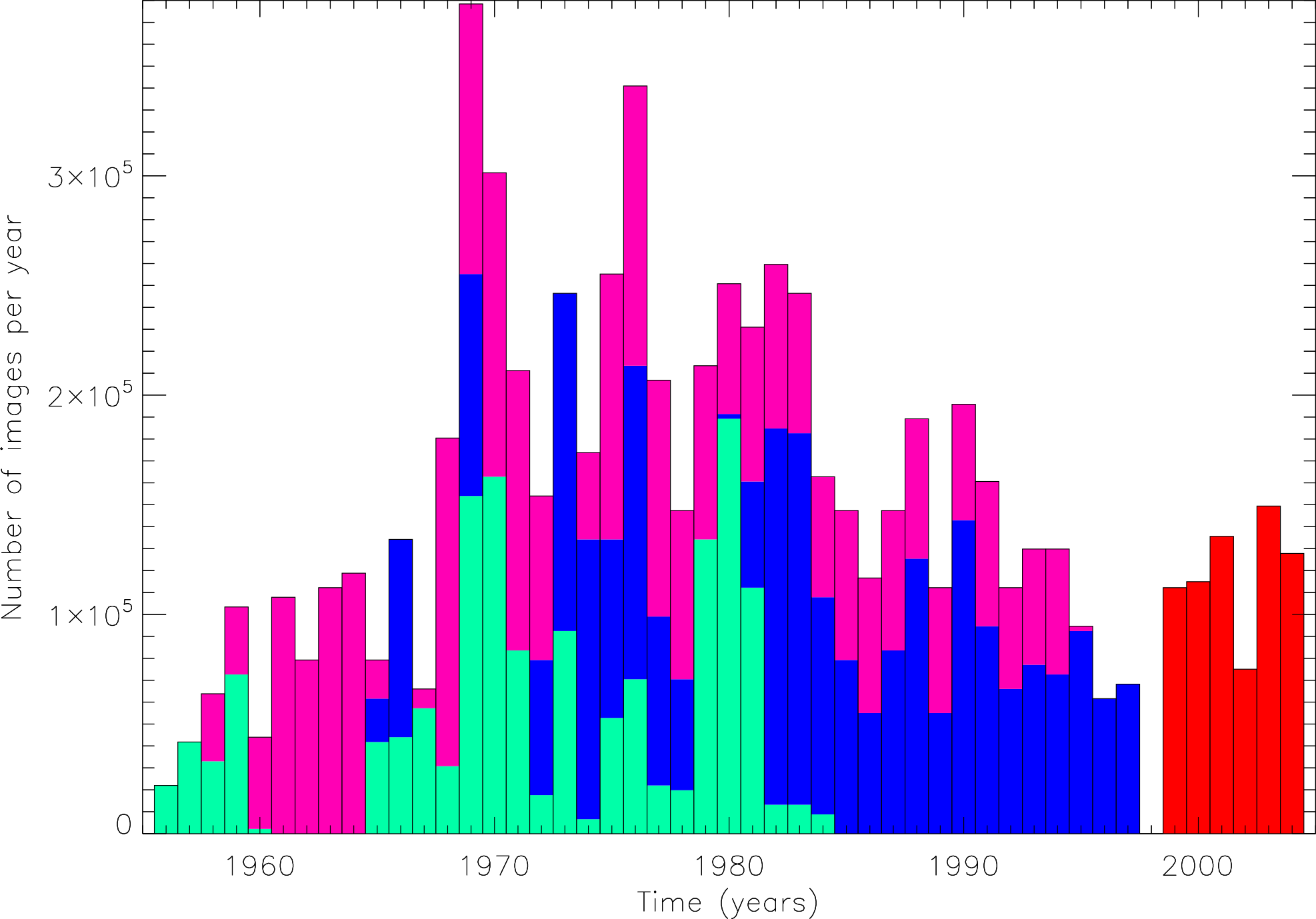}
 \caption[]{Number of H$\alpha$ images got with Meudon and OHP heliographs
 between 1956 and 2004 (green: Meudon, 11 mm images; pink:
 Haute Provence, 15 mm images; blue: Meudon, 3-wavelength (part Sun until 1985, full Sun after);
 red: Meudon, 3-wavelength CCD full Sun)
  } \label{ha4}
\end{figure}

More than 7 million images were produced both at Meudon and OHP
between 1956 and 2004 in the context of high cadence (60 seconds)
observations to investigate solar activity events (Table 1 and
Figure~\ref{ha4}).

\begin{figure}
\centering
\includegraphics[width=1.0\textwidth,clip=]{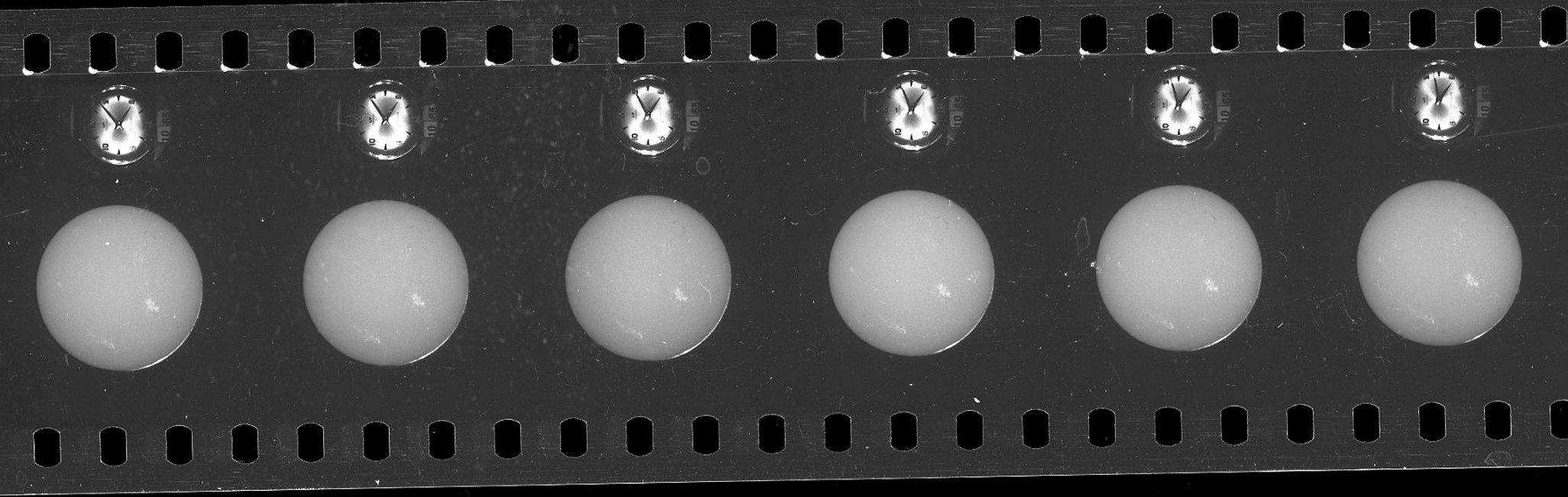}
 \caption[]{Example of 11 mm H$\alpha$ monochromatic full Sun images got with Meudon heliograph
 between 1965 and 1985, 0.75 {\AA} FWHM
 (4 October 1965 from 16:08 to 16:13 UT)
  } \label{ha1}
\end{figure}

The story began with the invention of the monochromatic birefringent
filter by \cite{lyot}. The first H$\alpha$ heliograph was built by
\cite{grenat}. Systematic observations started in 1956 with 35 mm
films (Figure~\ref{ha1}). The 45 meters long films (Kodak Technical
Pan TP2415) recorded each 2200 full Sun images, 15 mm diameter and
0.75 {\AA} FWHM. The film resolution (better than 125 lines/mm or
less than $1''$ at the Sun) was limited by the seeing ($2''$
typical). A similar instrument started at OHP in 1958 and worked
until 1995, producing 2.6 million images. The pair of sites allowed
an increase in temporal coverage, as clear sky at OHP is more
frequent. The 1957 IGY motivated the scientific program of both
instruments.

\begin{figure}
\centering
\includegraphics[width=1.0\textwidth,clip=]{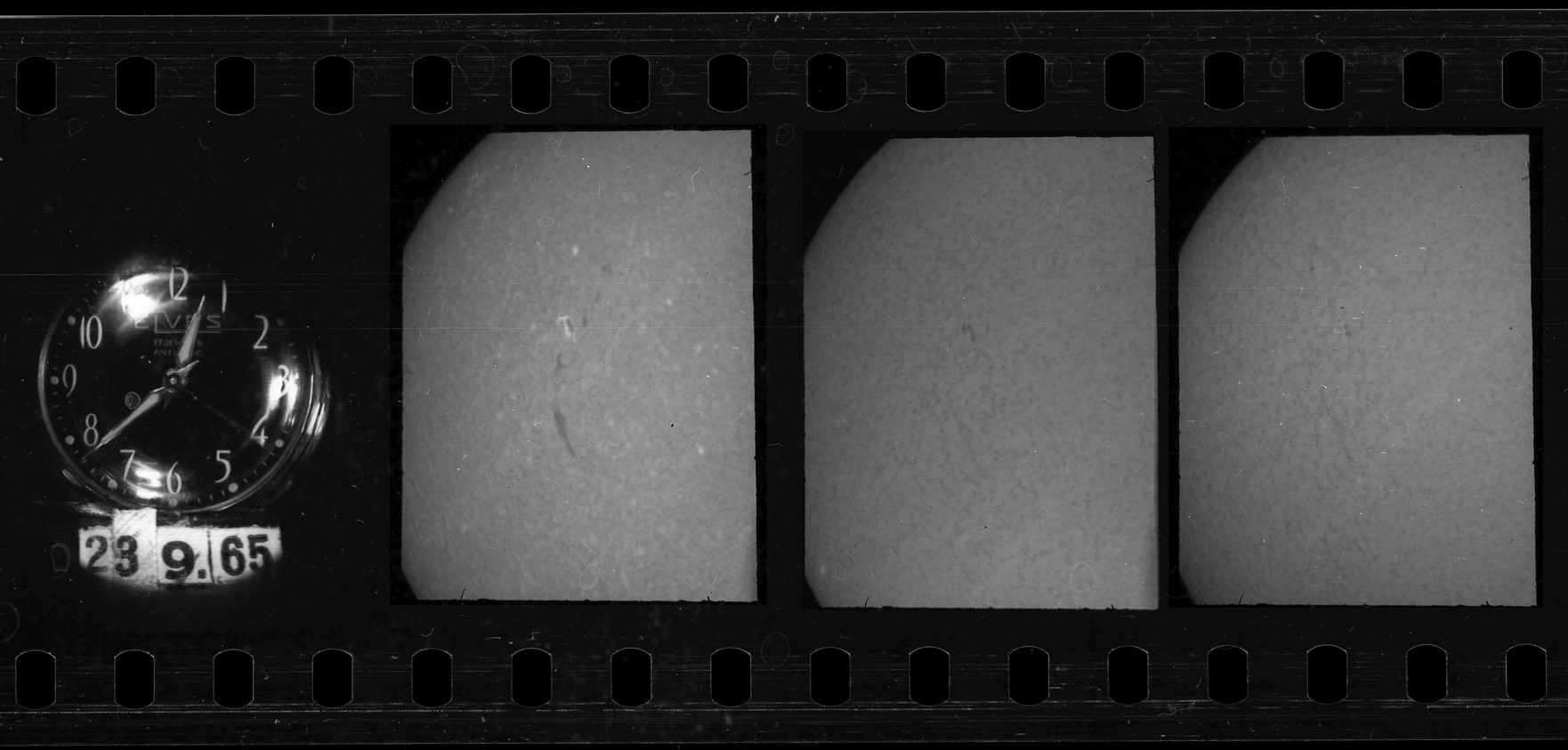}
 \caption[]{Example of 3-wavelength H$\alpha$ part Sun images got with Meudon heliograph
 between 1965 and 1984: line center, blue and red wings ($\pm$ 0.75 {\AA}), 0.75 {\AA}
 FWHM, here images of 4 October 1965 at 12:38 UT)
  } \label{ha2}
\end{figure}

Between 1960 and 1965, Meudon observations were interrupted but
continued at OHP. A new 3-wavelength instrument was built for
observations at H$\alpha$ center, blue and red wings ($\pm$ 0.75
{\AA}) using a 3-stage tunable Lyot filter with motorized rotating
plates (Figure~\ref{ha2}, \cite{Michard}, \cite{Demarcq}).
Observations started in 1965, covering about 30 \% of the solar area
(16 x 20 mm FOV selected in the 35 mm image of the Sun). For that
reason, a second routine, full disk and only at H$\alpha$ center,
started at the same time (11 mm solar diameter). Both instruments
had 0.75 {\AA} FWHM and were in production until 1984.

\begin{figure}
\centering
\includegraphics[width=1.0\textwidth,clip=]{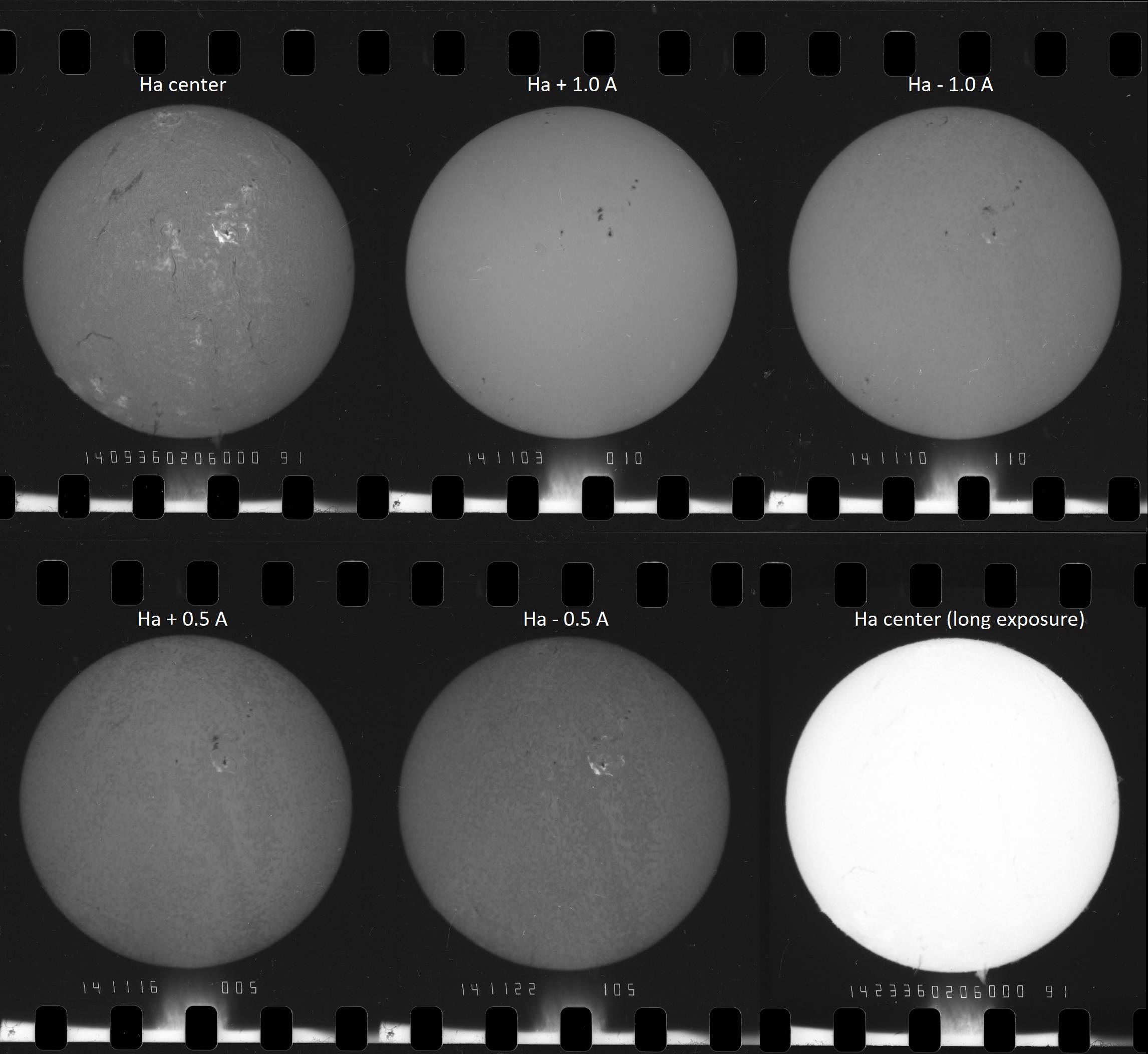}
 \caption[]{Example of 5-wavelength H$\alpha$ full Sun images got with Meudon heliograph
 between 1985 and 1997, line center, red and blue wings ($\pm$ 0.1 {\AA}, $\pm$ 0.50 {\AA})
 and overexposed line core for prominences (0.50 {\AA} FWHM,
 flare of 2 June 1991 at 14:11 UT).
  } \label{ha3}
\end{figure}

In 1985, a new tunable Lyot filter, with better performance
(H$\alpha$ center, blue and red wings ($\pm$ 0.50 {\AA}, $\pm$ 1.0
{\AA}, 0.50 {\AA} FWHM) was developed (Figure~\ref{ha3},
\cite{Demarcq}). It was mounted between two lenses of 360 mm focal
length; this afocal system was fed by a 150/2250 mm telescope.
Observers chose 3 or 5 wavelengths according to solar activity
level, plus a long exposure image for prominences at line center.
This 5-stage Lyot filter (11 {\AA} distance between maxima) was
optimized to cut secondary lobes. The H$\alpha$ peak was isolated by
a three cavity 3.5 {\AA} FWHM blocking filter. This device provided
full disk images (21 mm diameter), such that the old 11 mm H$\alpha$
center routine was stopped. A 1536 x 1024 cooled CCD camera from
Princeton Instruments, supporting a 140 mm objective, replaced the
film in 1999 (9 $\mu$ pixel, 12 bits, cooled KAF1600 sensor),
producing 0.7 million FITS images until 2004 (Table 1 shows that,
contrarily to the film, CCD images were undersampled). FITS data are
off line (CD/DVD), but freely available upon request. Light curves
and quick look MPEG movies are on line at the BASS2000 solar
database (\url{http://bass2000.obspm.fr/home.php?lang=en}).

\begin{table}
\begin{tabular}{cccccc}
  \hline
  Site & Date & Wavelength & Sun diameter & Resolution & FWHM  \\
       &      & number     &   [mm]       & [arcsec] & {\AA} \\
  \hline
  Meudon (films) & 1956-1960 & 1 & 15 (FS) & 1.0 & 0.75  \\
                 & 1965-1982 & 1 & 11 (FS) & 1.4 & 0.75  \\
                 & 1965-1984 & 3 & 35 (PS) & 0.5 & 0.75  \\
                 & 1983-1984 & 1 & 16 (FS) & 1.0 & 0.75  \\
                 & 1985-1997 & 3 or 5 & 21 (FS) & 0.8 & 0.50  \\
  Meudon (CCD)   & 1999-2004 & 3 &  8.5 (FS) & 3.7 & 0.50  \\
  OHP (films)    & 1958-1995 & 1 & 15 (FS) & 1.0 & 0.75  \\

  \hline
\end{tabular}
\caption{1956-2004 H$\alpha$ observations at Meudon and OHP (FS =
full Sun; PS = part Sun). Three wavelengths mean that line center,
blue and red wings ($\pm 0.75$ \AA) are observed in sequence, while
for five wavelengths, we have four wing positions ($\pm 1.0$ \AA,
$\pm 0.5$ \AA).}
\end{table}

Figure~\ref{ha4} summarizes observations made between 1956 and 2004.
About 6.5 million images have been recorded on 35 mm films (a total
of 130 kilometers), and 0.7 million of CCD images are archived. The
detailed list is available at:

\url{http://www.lesia.obspm.fr/perso/jean-marie-malherbe/heliograph/index.html}

The cost makes impossible a systematic scan of the 3000 films of
this exceptional collection, but films of interest can be digitized
individually upon request (8 bits, 157 pixels/mm scans, sampling
better than $1''$/pixel, TIF format, one file for 100 mm of film).

Observations have been mainly used by the Meudon group and exploited
qualitatively, as long as films were not digitized, to investigate
chromospheric flares and filament instabilities. The
multi-wavelength filters allowed to study mass motions in active
regions.

Concerning flares, \cite{Martres3} found a relationship with radio
bursts observed in Nan\c{c}ay. \cite{Mouradian1} suggested that a
flare is composed of several elementary eruptive phenomena in
relation with magnetic emerging flux, involving the presence of both
cold (the surging arch at $10^{4}$ K) and hot (the flaring arch at
$10^{7}$ K) magnetic loops. \cite{Martres1} found evidence for
homologous flaring. \cite{Mouradian2} compared H$\alpha$ features
and X-ray brigtenings at flare onset. Later with the digital version
of the 3-wavelength instrument, \cite{Pick} and \cite{Maia}
investigated CMEs and flares using H$\alpha$ and radio data.

As for filaments, \cite{Mouradian4} discovered the existence of
rigid rotation points (the pivot points) in some filaments which
could play a role in instabilities. \cite{Soru} suggested that
heating and cooling mechanisms could explain several cases of sudden
disappearances and reappearances. Hence, \cite{Mouradian3} proposed
two physical classes of disappearences, involving either thermal or
dynamic processes.


\section{The New Heliograph} \label{S-New-Version}

A small rolling house has been built at Calern observatory (1270 m
elevation) in order to have better weather and seeing conditions
than at Meudon or OHP. The equatorial mount was constructed by
Valmeca. It supports the instruments which are enclosed in a 1.7
$\times$ 0.5 $\times$ 0.5 m$^{3}$ box, thermally regulated above
ambient temperature at $27^{\circ}$ C by active heating and passive
cooling.

\begin{figure}
\centering
\includegraphics[width=1.0\textwidth,clip=]{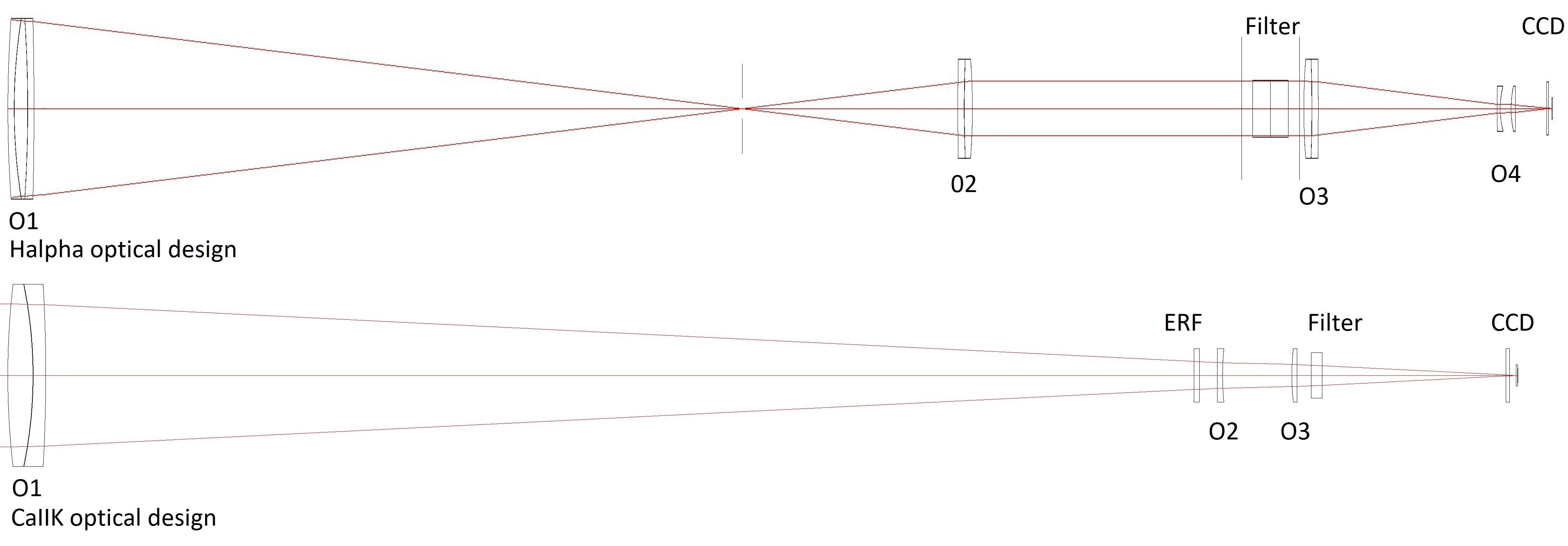}
 \caption[]{Optical design of the two H$\alpha$ (top) and
 the Ca\textsc{ii} K (bottom) telescopes.} \label{design}
\end{figure}

The new instrument is composed of three telescopes (Tables 2 and 3)
corresponding to the design of Figure~\ref{design}. The two
H$\alpha$ telescopes are identical, except for the Fabry P\'{e}rot
etalons manufactured by DayStar corporation (professional series).
The first filter is line core centered, in order to observe active
regions, flares and filaments. The second one is adjusted to the
blue wing (choice discussed in Section 5). The third telescope is
centered on Ca\textsc{ii} K with an interference filter from Barr
company. The equivalent focal length is 983 mm providing a 9.13 mm
solar image and $35'$ FOV. Each detector (Table 4) is a 12 bits
cooled camera from Quantum Scientific Imaging using shutterless
interline CCD sensors from Sony (ICX694 and 814 respectively for
H$\alpha$ and Ca\textsc{ii} K). The readout noise is 7 electrons RMS
at 8 MHz. Exposure times are shorter than 10 ms.

H$\alpha$ telescopes are afocal systems. O1 is the entrance
objective (Takahashi TSA102, 102 mm diameter) protected in full
aperture by a Baader energy rejection filter (ERF). The Fabry
P\'{e}rot etalon is located in the pupil image at F/30 between O2
and O3 (maximum beam aperture constrained by the manufacturer). O4
(a set of two separate lenses) is an amplifier and field corrector.
A 80 mm diaphram limits the optical resolution to $2.0''$.

\begin{figure}
\centering
\includegraphics[width=1.0\textwidth,clip=]{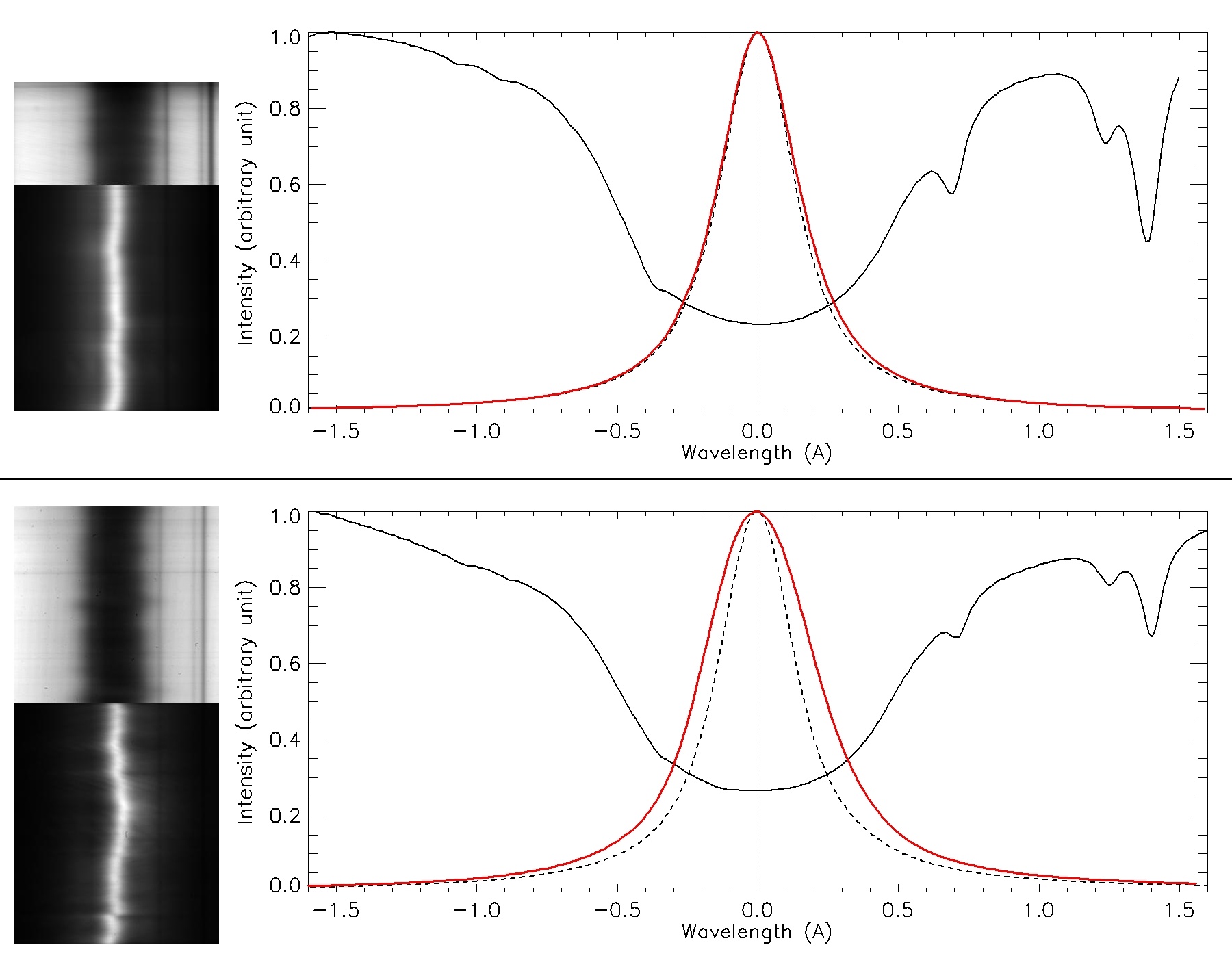}
 \caption[]{Calibration at F/60 of the two H$\alpha$ filters (0.34 {\AA} FWHM top,
 0.46 {\AA} FWHM bottom). Left: H$\alpha$ observed line at disk center
 with or without the filter. Right: transmission functions; black line: observed
 line profile; dashed line:
 local typical transmission (0.32 {\AA} FWHM); red line: global transmission averaged over
 the filter surface for pupil plane application.} \label{calib}
\end{figure}

The calibration of the two H$\alpha$ filters was done at F/60 with
the high resolution spectrograph (R = 300000) of the Meudon solar
tower. The surface filter (31.75 mm diameter) was scanned by the
slit. Both filters exhibit 0.32 {\AA} FWHM (homogeneous over the
filter), but the scans show that the CWL varies from place to place.
The pupil plane location of the afocal design allows to correct this
effect and delivers a uniform CWL image, but, as a counterpart, it
increases the global FWHM (Figure~\ref{calib}). Hence, filter 1
(2019) is almost perfect (0.34 {\AA} global FWHM) and is dedicated
to H$\alpha$ center images. Filter 2 (2009) has some optical defects
(0.46 {\AA} global FWHM) and is used for H$\alpha$ wing (Moreton
waves detection); it exhibits a 10\% parasitic internal reflection
which vanishes in the running difference process (Section 5). Each
Fabry P\'{e}rot etalon is solid (mica spaced) and $\lambda$/10
surfaced. The H$\alpha$ peak is selected by an internal 2-cavity
8-10 {\AA} FWHM blocking filter. Adjacent peaks are about 20 {\AA}
apart with only 0.34 \% transmission of the central peak. However,
the blocking filter of etalon 2 is shifted by 2.8 {\AA}. A colored
longpass glass filter (RG630) suppresses short wavelengths (UV,
blue, green), while IR radiation is rejected by a shortpass filter.

\begin{figure}
\centering
\includegraphics[width=1.0\textwidth,clip=]{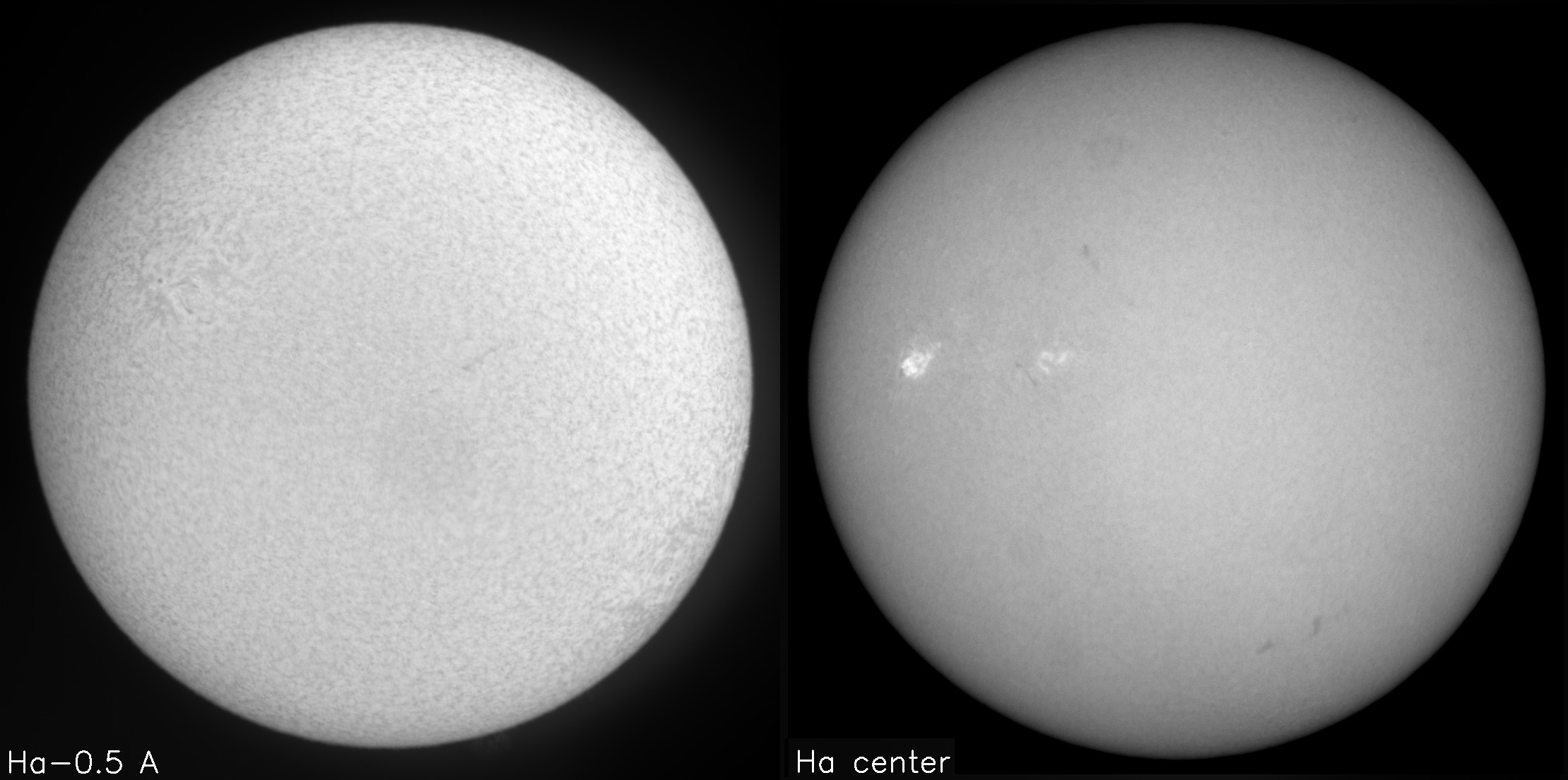}
 \caption[]{Images obtained with the 0.46 {\AA} FWHM filter (29 August 2017)
 in H$\alpha$-0.5 {\AA} (blue wing, left) and with the 0.34 {\AA} FWHM filter (21 March 2019)
 in H$\alpha$ line core (right).} \label{images}
\end{figure}

Both etalons are thermally regulated; the CWL is temperature
dependant ($9^{\circ}$C/{\AA} typical). As the wavelength shift is
very slow (about 0.1 {\AA}/minute), contrarily to Lyot tunable
filters based on rotating plates, it is not possible to scan line
profiles, so that two filters are required for two wavelength
positions (Figure~\ref{images}).

The Ca\textsc{ii} K telescope provides a magnetic field proxy for
active regions and faculae \citep{Pev}. As the filter is broad (1.5
{\AA} FWHM), the design does not need to be afocal. O1 is the
entrance objective (Takahashi FS102, 102 mm diameter). O2 and O3
constitute a focal amplifier. The 3934 {\AA} filter is located close
to the image plane and protected by an ERF. A 80 mm diaphragm limits
the resolution to $1.25''$.

Meteospace is an autonomous station: opening and closing the dome,
weather control, catching the Sun, observations, data processing,
real-time dissemination, database archiving, all operations are
automated. Sensors permanently control the instrument and can decide
to stop and close in case of cloud, rain, alarm or failure; alerts
are sent to the local staff by a Short Message Service (SMS). The
entry point to the data is the BASS2000 service, even if the full
archive is physically located at Nice.

All data will be delivered freely to the solar community. Raw (level
0) data as well as standard processed data (level 1) will be
available in real-time JPEG for quick look purpose and FITS for
scientific use. Level 1 includes corrections as dark current,
distorsion and solar image rotation to present solar north up. The
observing cadence is, respectively for H$\alpha$ and Ca\textsc{ii}
K, 15 and 60 seconds. The system will run systematically (weather
permitted) under seeing conditions better than $2''$ (sampling
reported in Table 4). Following \cite{Thompson}, FITS headers
include World Coordinates System (WCS) informations concerning image
pointing and orientation, allowing the use of positioning tools.

\begin{table}
\begin{tabular}{lcccccc}
  \hline
  Telescope & O1   & O2   & O3   & O4   & $f_{equiv}$ & FOV \\
            & [mm] & [mm] & [mm] & [mm] & [mm]        & [arcmin]      \\
   \hline
  H$\alpha$       & 816      & 250         & 250 & -50/+60 & 983 & 35 \\
  Ca\textsc{ii} K & 820      & -220        & 250 &         & 983 & 35 \\

  \hline
\end{tabular}
\caption{Optical design (the two H$\alpha$ telescopes are identical,
apart the filter).}
\end{table}

\begin{table}
\begin{tabular}{ccccccc}
  \hline
  Line & Wavelength & FWHM  & Temperature & Peak     & max. & Blocking \\
       &  (CWL)     & (global) &             & distance & finesse & filter FWHM \\
       & [{\AA}]    &[{\AA}]& [$^{\circ}$C] & [{\AA}]&         & [{\AA}]\\
  \hline
  H$\alpha$ 1        & 6562.8 & 0.34 & 38 & 24.7 & 77 & 10.0\\
  H$\alpha$ 2        & 6562.3 & 0.46 & 65 & 19.7 & 61 & 8.6 \\
  Ca\textsc{ii} K    & 3933.7 & 1.5  & 23 &  &  &  \\
  \hline
\end{tabular}
\caption{Filter characteristics.}
\end{table}

\begin{table}
\begin{tabular}{cccccccc}
  \hline
  Line & Dynamic & Pixel   & Sampling & format & Quantum   & Full  & Gain\\
       & range   &         & pixel    &        & efficiency& well  &    \\
       &         &[microns]&[arcsec]& [pixels] & [\%]      &[\={e}]&[\={e}/ADU]\\
  \hline
  H$\alpha$       & 2700 & 4.54 & 0.96 & 2758$\times$2208 & 65 & 19000 & 0.40 \\
  Ca\textsc{ii} K & 2570 & 3.69 & 0.78 & 3388$\times$2712 & 60 & 18000 & 0.36 \\
  \hline
\end{tabular}
\caption{Detector characteristics.}
\end{table}

\section{New Perspectives for Flare Onset and Moreton Wave Detection} \label{S-Moreton}

The purpose of the two high cadence H$\alpha$ telescopes is to
improve the detection of flare onset and Moreton waves. Such events
are suspected to be the counterpart of fast coronal EUV waves in the
chromosphere, which could be compressed by the coronal shock above,
producing 5-10 km s$^{-1}$ downflows \citep{Warmuth2015}.

\begin{figure}
\centering
\includegraphics[width=1.0\textwidth,clip=]{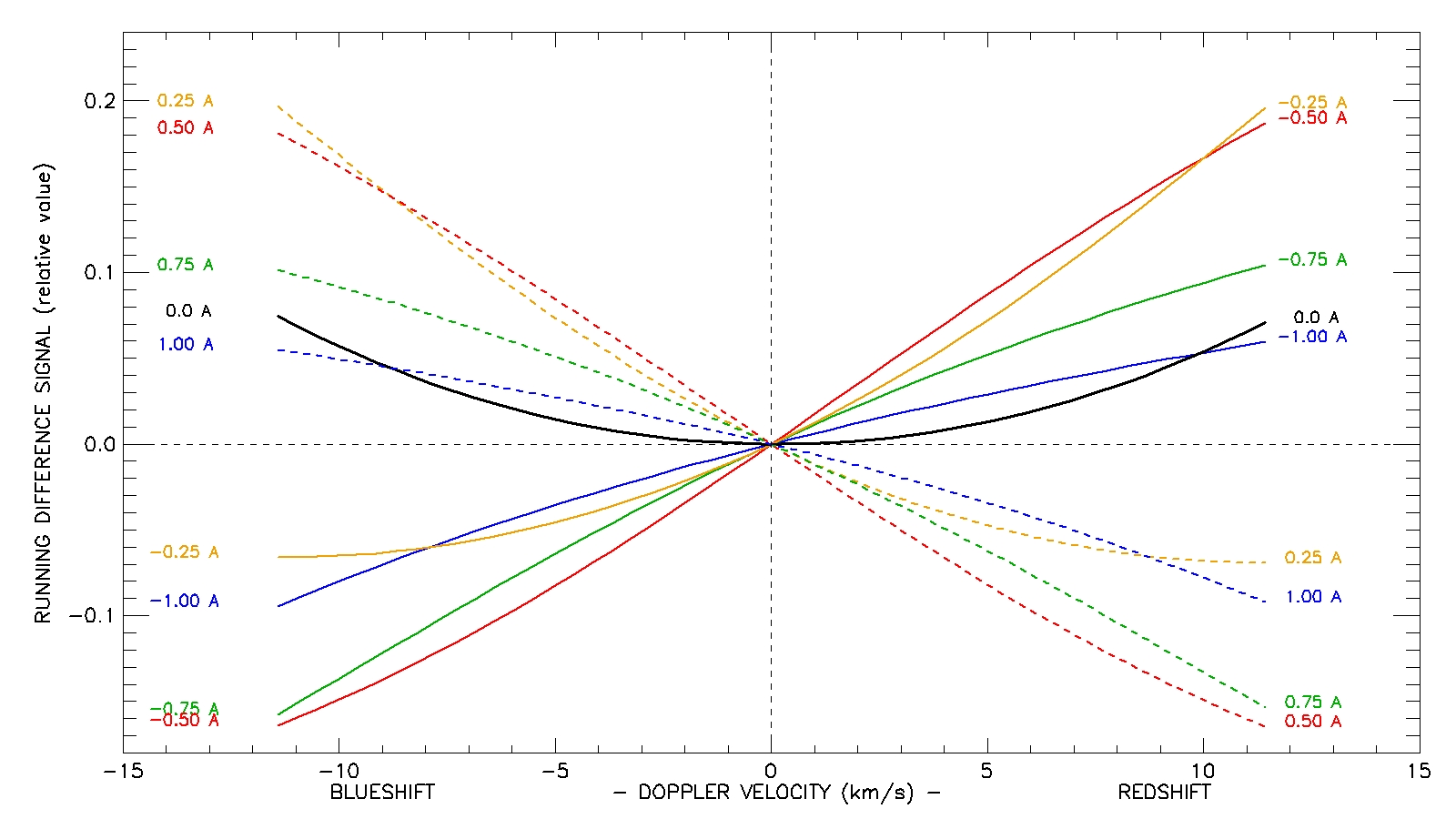}
 \caption[]{Running-difference signal as a function of Doppler velocity for different
 shifts of the filter bandpass (-1.0 to +1.0 {\AA} by 0.25 {\AA} step). Solid lines: blueshifted bandpass;
 dashed lines: redshifted bandpass. Bandpass shift: no shift (black), $\pm 0.25$ {\AA} (yellow),
 $\pm 0.50$ {\AA} (red),$\pm 0.75$ {\AA} (green),$\pm 1.0$ {\AA} (blue).} \label{rundif}
\end{figure}

We simulated the detection of Moreton waves by the 0.46 {\AA} FWHM
filter using the difference signal between a shifted profile and the
quiet sun profile (Figure~\ref{rundif}). We used the atlas profile
from \cite{del} altered by 15 \% scattered light and shifted from
-10 to +10 km s$^{-1}$. The filter CWL was moved from the blue to
the red wing (-1.0 {\AA} to +1.0 {\AA} by 0.25 {\AA} step). The best
sensitivity is obtained, for chromospheric downflows, when the
filter selects the blue wing (-0.5 {\AA}, 9 \% and 17 \% enhancement
respectively for 5 and 10 km s$^{-1}$). In the case of simultaneous
up or downflows, both wings are convenient. These results are
illustrated by movie 1 showing running-differences of the 28 October
2003 event observed by the tunable Meudon Lyot filter (H$\alpha$
center and wings).

\begin{figure}
\centering
\includegraphics[width=1.0\textwidth,clip=]{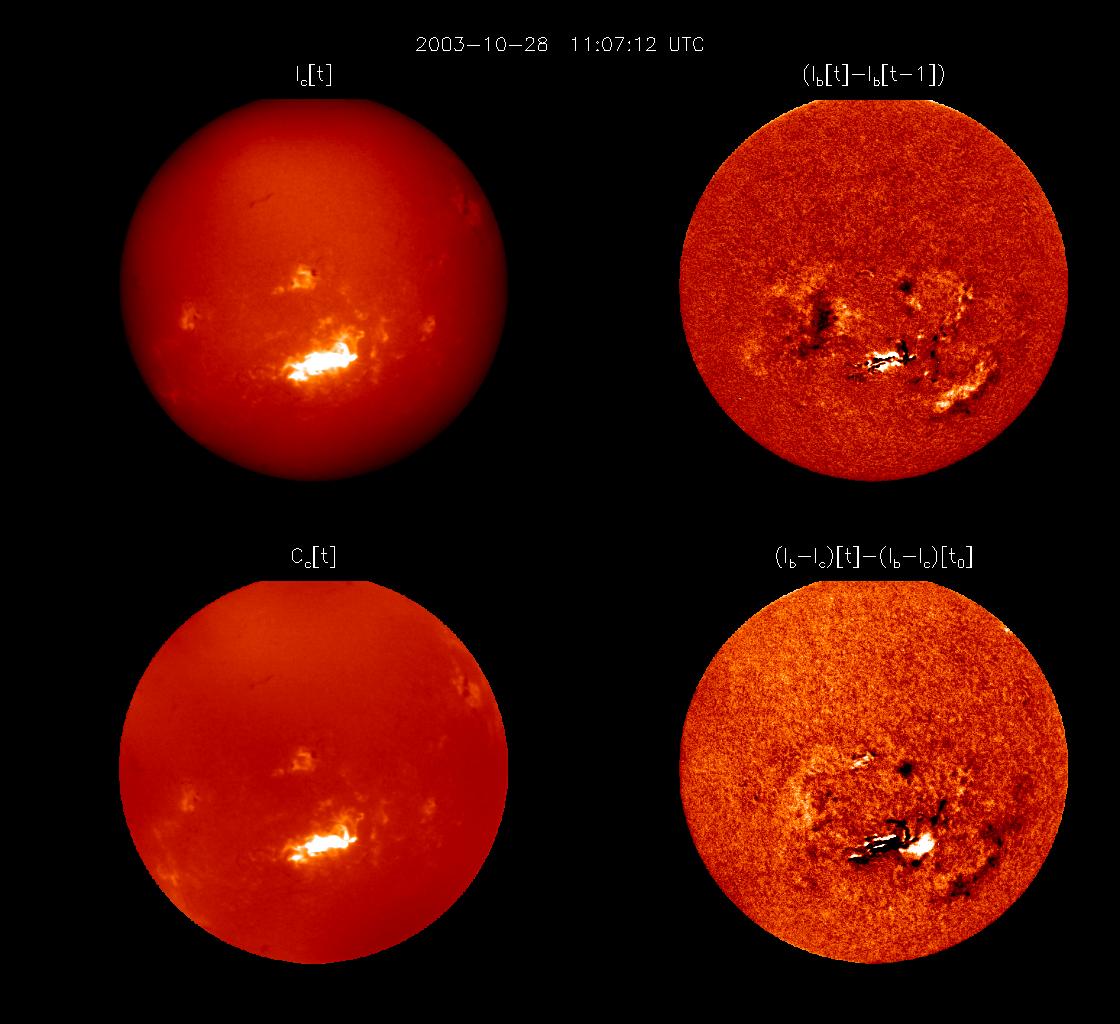}
 \caption[]{Simulation (based on the 28 October 2003 event at 11:07:12 UT) of
 data which will be produced by the new heliograph. Top: real-time images; central
 intensity (left) and running-difference of the blue wing intensity (right). Bottom:
 post-processed images; contrast at line center suppressing the limb darkening (left)
 and Doppler proxy (right) issued from the intensity difference between
 the blue wing and line center, after correction by an offset observed before the event.
 } \label{methods}
\end{figure}

As Moreton waves will probably not occur before the next maximum, a
simulation of typical data which will be provided by the new
instrument, based on 28 October 2003 data, is displayed in
Figure~\ref{methods}. Real-time images and movies will be produced,
as central intensities $I_{c}(t)$ or contrasts $C_{c}(t)$ and
running-differences of the blue wing intensity $I_{b}(t)-I_{b}(t-1)$
or contrast $C_{b}(t)-C_{b}(t-1)$. Contrasts give smoother results
and are defined by $C = \frac{I}{LD} - 1$, where LD is the limb
darkening function. It is built from the image by taking median
values in concentric rings.

Post-processed images will also be delivered. The two H$\alpha$
filters can be combined to give $I'(t)= I_{b}(t)-I_{c}(t)$, $I_{b}$
and $I_{c}$ beeing respectively the blue wing and central intensity.
Then, an offset $I'(t_{0})$, measured before the event, is
subtracted to the current signal $I'(t)$, as suggested by
\cite{muhr}. This provides a proxy of the Doppler velocity. Results
issued from both wings would be better, but this cannot be done with
our instrument. The method can be applied to contrasts instead of
intensities, as shown by movie 2.

Hence, efforts to implement real-time detection of Moreton events
using running differences of blue wing intensities are underway and
will be refined using initial datasets from the instrument as it is
commisionned. This tool, together with the survey of flare
brightening in line core, could be used to monitor flare onset and
anticipate their possible impact at the Earth.

\section{Discussion and Conclusion} \label{S-Conclusion}

\begin{figure}
\centering
\includegraphics[width=1.0\textwidth,clip=]{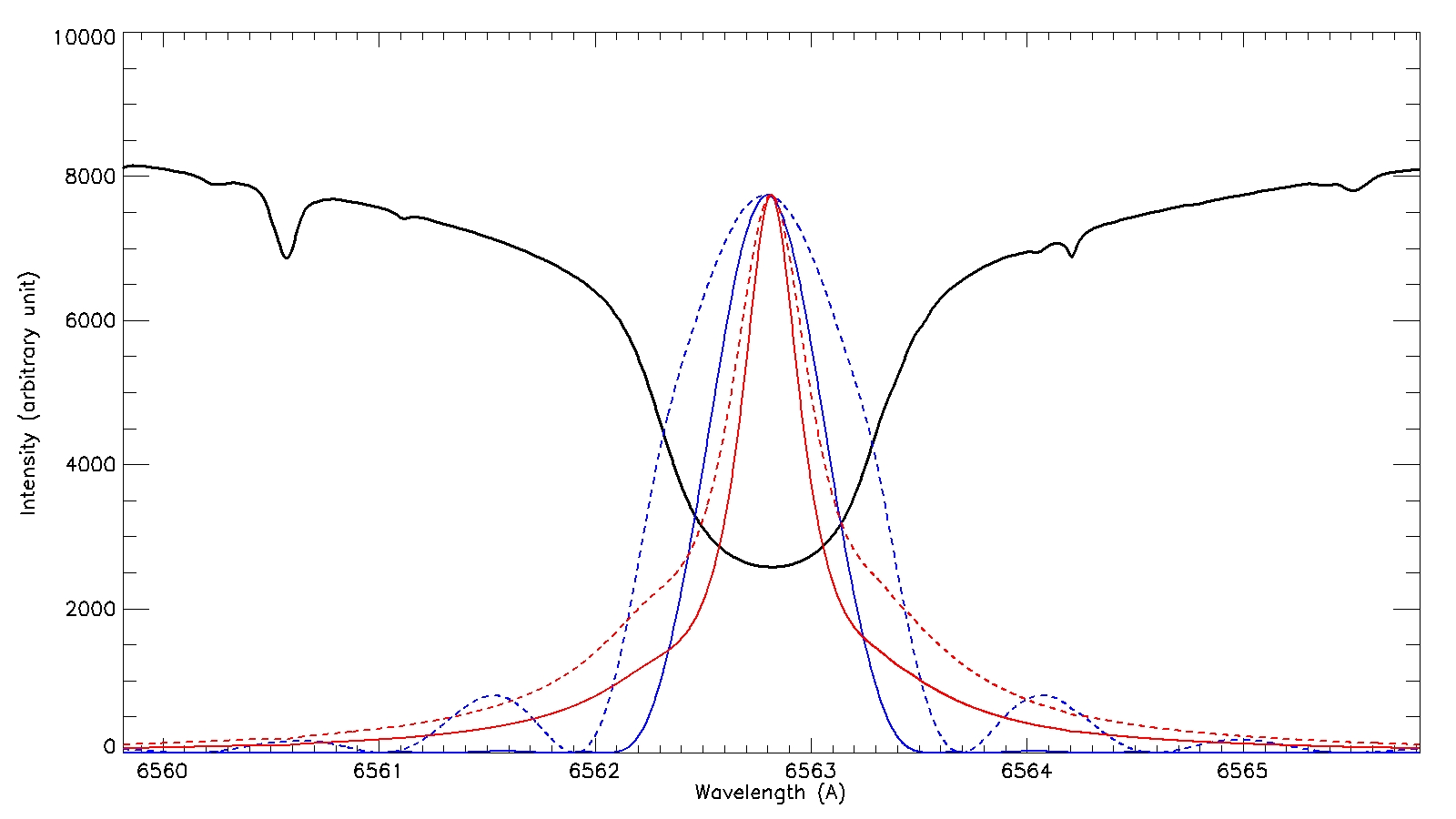}
 \caption[]{Spectral energy transmitted by old H$\alpha$ Meudon Lyot prototypes
 (blue, 1985-2004 solid, 1965-1984 dashed) and the two DayStar Fabry P\'{e}rot etalons
 (red, solid and dashed). } \label{compar}
\end{figure}

More than 7 million images at 60 seconds cadence (flares, CME onset,
filament instabilities, Moreton waves) have been obtained at Meudon
and OHP from 1956 to 2004 using H$\alpha$ Lyot filters, either line
centered or 3-wavelength (center + wings). Systematic observations
will restart in 2020 at Calern observatory using more compact and
cheaper technology. The new heliograph is based on two commercial
Fabry P\'{e}rot etalons. However, such filters have far photospheric
wings (Lorentzian shape), contrarily to Lyot filters which mainly
pass the chromosphere (Figure~\ref{compar}). Let us define the
energy bandpass (EB) as the wavelength interval containing half of
the energy transmitted by the filter. It is computed from the
spectral energy (the product of the filter transmission curve by the
quiet H$\alpha$ disk center line profile). The EB of the Meudon
3-wavelength Lyot filters (1965-1984 and 1985-1997), respectively
0.62 {\AA} and 0.34 {\AA}, is always smaller than their FWHM (the
second filter was optically more optimized). In comparison, the
distance between inflexion points of the H$\alpha$ line is 1.0
{\AA}. On the contrary, our industry Fabry P\'{e}rot devices provide
0.53 {\AA} and 0.77 {\AA} EB, greater than their respective FWHM.
Hence, more photospheric light will pass in comparison to previous
Lyot prototypes, reducing for instance filament contrasts. However,
adding 2.0 or 3.0 {\AA} FWHM pre-filters would allow, in the future,
to minimize photospheric wing contributions and achieve former
performance.

 Three telescopes are available:
high cadence (15 seconds) H$\alpha$ center and blue wing, and medium
cadence (60 s) Ca\textsc{ii} K (active regions and magnetic field
proxy). Real-time and post-processed images for scientific purpose
will be produced: H$\alpha$ center for filament eruption and flare
brightening detection, as well as running-difference of blue wing
for Moreton waves (often involved in large flares). All images
(quick look JPEG and scientific FITS) will be freely available to
the solar community through BASS2000, without any delay.

After 16 years of interruption (due to the lack of observers), this
new automatic instrument will resume the exceptional survey started
at the IGY. It will complete the GHN and GONG networks. GONG has
only one european station in Tenerife. Weather conditions in
C\^{o}te d'Azur are the best available in France for continuous
observations. This opportunity allows to offer more time coverage to
existing networks, improved cadence, core and wing observations. The
new routine is dedicated to space weather scientific studies related
to flares, CME and filament instabilities, as well as operational
monitoring and forecast by the french Air Force. Detailed
opto-mechanical drawings are available for reproduction at other
places.

\begin{acks}

We thank the anonymous referee for helpful comments and suggestions.
We are indebted to the Meteospace technical team: G. Barbary, C.
Blanchard, I. Bual\'{e}, S. Cnudde, C. Collin, C. Colon, D.
Crussaire, A. Demathieu, C. Imad, Ph. Laporte, R. Lecocguen, M.
Ortiz, Ch. Reni\'{e}, D. Ziegler (Observatoire de Paris) and Y.
Bresson, F. Guitton, F. Morand, C. Renaud (Observatoire de la
C\^{o}te d'Azur). We are also grateful for financial support to the
Direction G\'{e}n\'{e}rale de l'Armement, the scientific councils of
Paris and Nice Observatories, the IDEX UCA/JEDI acad\'{e}mie 3, Ile
de France r\'{e}gional council and the Programme National Soleil
Terre (INSU/CNRS). K. Dalmasse is supported by the Centre National
d'Etudes Spatiales (CNES).
\end{acks}

\section*{Disclosure of Potential Conflicts of Interest}

The authors declare that they have no conflicts of interest.

%
\appendix

Electronic Supplemental Material (movies in MPEG 4 format).

\begin{description}

\item[\textnormal{i)}] Movie 1: Running-difference of contrasts applied to the typical
Moreton event of 28 October 2003 from 10:41 UT to 11:22 UT (Meudon
H$\alpha$ heliograph). Contrasts C are derived from intensities I
after limb darkening (LD) correction ($C = \frac{I}{LD} - 1$). Time
step 60 seconds. Top left: contrasts at line center; top right:
running-differences of line center contrasts; bottom:
running-differences of blue wing contrasts (H$\alpha$ - 0.5 {\AA},
left) and red wing contrasts (H$\alpha$ + 0.5 {\AA}, right).

\item[\textnormal{ii)}] Movie 2: simulation of typical images and movies which will be provided
by the new instrument, based on 28 October 2003 data, from 10:41 UT
to 11:22 UT (Meudon heliograph). Top: real-time processing will
provide running-difference of contrasts (left) and intensities
(right) in the blue wing (H$\alpha$ - 0.5 {\AA}). Contrasts C are
derived from intensities I after limb darkening (LD) correction ($C
= \frac{I}{LD} - 1$). Bottom: post-processing based on \cite{muhr}
method: base-difference between contrasts (left) or intensities
(right) at time $t$ and an offset measured before the event
(reference time $t_{0}$), combining blue wing and line core images
of the two H$\alpha$ telescopes.

\end{description}

\bibliographystyle{spr-mp-sola}
\bibliography{JMM_sola_V3}

\IfFileExists{\jobname.bbl}{} {\typeout{}
\typeout{****************************************************}
\typeout{****************************************************}
\typeout{** Please run "bibtex \jobname" to obtain} \typeout{**
the bibliography and then re-run LaTeX} \typeout{** twice to fix
the references !}
\typeout{****************************************************}
\typeout{****************************************************}
\typeout{}}

\end{article}

\end{document}